\documentclass[final]{IEEEtran}
\IEEEoverridecommandlockouts

\usepackage{cite}
\usepackage{amsmath,amssymb,amsfonts}
\usepackage{algorithmic}
\usepackage{graphicx}
\usepackage{textcomp}
\usepackage{xcolor}
\usepackage{booktabs}
\usepackage{url}
\usepackage{amssymb}   
\usepackage{pifont}
\newcommand{\xmark}{\ding{55}}
\def\BibTeX{{\rm B\kern-.05em{\sc i\kern-.025em b}\kern-.08em
    T\kern-.1667em\lower.7ex\hbox{E}\kern-.125emX}}
\begin{document}

\title{\textsc{Vhdl-RepoBench}: A Repository-Level Benchmark for Evaluating Large Language Models on VHDL Design Generation}

\author{
\IEEEauthorblockN{
Prashanth Vijayaraghavan\IEEEauthorrefmark{1},
Akul Malhotra\IEEEauthorrefmark{1},
Ashutosh Jadhav,
Ehsan Degan,
Vandana Mukherjee
}

\IEEEauthorblockA{
\textit{IBM Research}\\
San Jose, CA, USA\\
\{prashanthv,Akul.Malhotra\}@ibm.com,
\{ashutosh,edehgha,vandana\}@us.ibm.com,
}

\thanks{\IEEEauthorrefmark{1}These authors contributed equally to this work.}
}

\maketitle

\begin{abstract}
Large Language Models (LLMs) are increasingly applied in hardware design automation, demonstrating strong potential in generating and understanding hardware description languages. However, most existing benchmarks focus on Verilog, with limited evaluation of VHDL, which remains widely used in industry and academia for FPGA and safety-critical systems. To address this gap, we introduce \textsc{Vhdl-RepoBench}, a large-scale, cross-file, repository-level benchmark for assessing LLM capabilities on realistic VHDL design generation and analysis tasks. \textsc{Vhdl-RepoBench} curates $\sim$100 open-source VHDL repositories, encompassing $\sim$2.5k VHDL files and $\sim$500 testbenches, and provides structured problem statements, module stubs, and self-verifying testbenches. The benchmark enables comprehensive evaluation across syntax, semantic correctness, hierarchical reasoning, cross-file dependency resolution, and functional verification. We evaluate several state-of-the-art models, including GPT-4o, Llama-3-70B, Qwen2.5-72B, CodeLlama-70B, and multi-step reasoning approaches such as Reflexion and CoDes. Results reveal that while current LLMs achieve moderate line- and block-level accuracy, substantial challenges remain in multi-file reasoning, hierarchical design understanding, and specification-to-module generation. \textsc{Vhdl-RepoBench} represents the first large-scale VHDL-focused benchmark and provides a valuable resource for the hardware design community to evaluate, compare, and advance LLM capabilities for practical VHDL development.
\end{abstract}


\section{Introduction}
Large language models (LLMs) are rapidly transforming hardware design automation by enabling automated generation, completion, and reasoning over Register-Transfer Level (RTL) code \cite{yu2025spec2rtl,dave,chipchat,mgverilog,akyash2025rtl++, zhao2024codev, zuo2025complexvcoder}. While most recent progress has focused on Verilog, VHDL remains widely used across industry, research, aerospace, and safety-critical domains. Compared to Verilog, VHDL enforces a richer type system, stricter modular structure, explicit package usage, and well-defined generic configurations. These features make VHDL more verbose and semantically constrained, demanding deeper structural and cross-file reasoning from LLMs.
\begin{figure}[t!]
    \centering
    \includegraphics[width=0.48\textwidth]{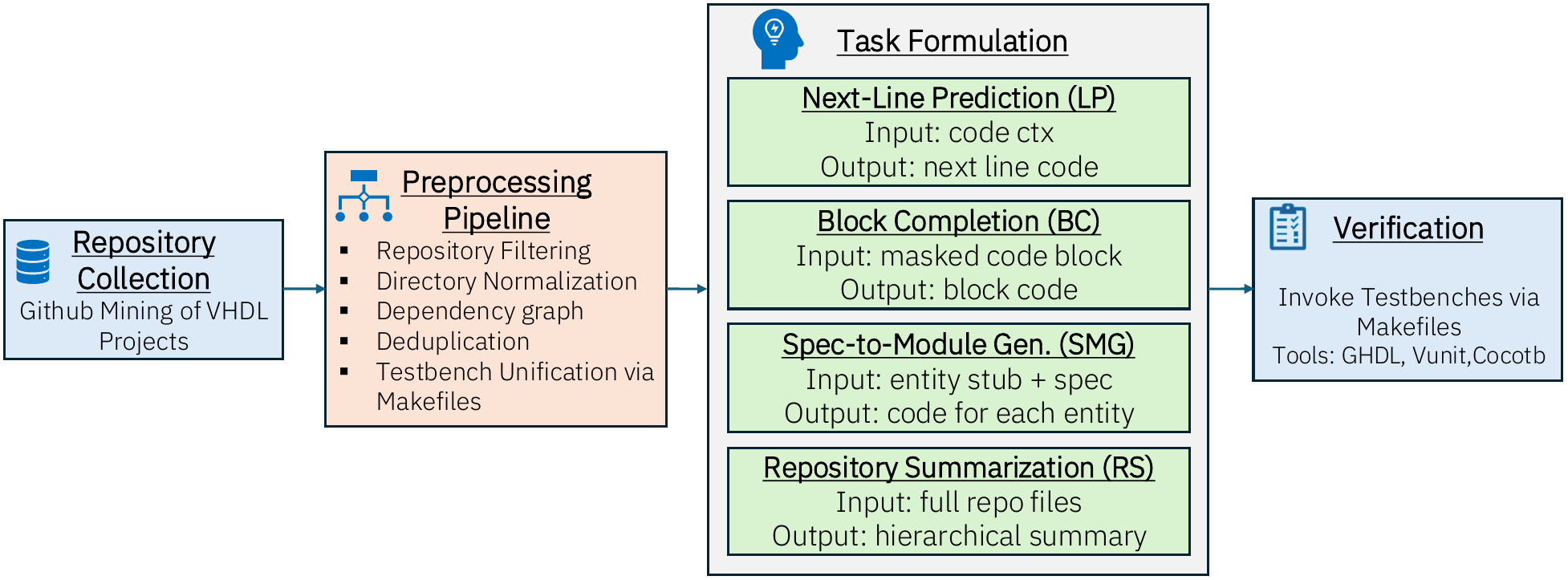}
    \caption{Overview of \textsc{Vhdl-RepoBench}. Public GitHub repositories are processed via a preprocessing pipeline to construct the dataset. The benchmark supports 4 tasks: Next-Line Prediction (LP), Block Completion (BC), Specification-to-Module Generation (SMG), \& Repository Summarization (RS). Verification module enables evaluation of LLM performance on multi-file VHDL reasoning \& generation.}
    \label{fig:vhdl_overview}
\end{figure}

Despite the prevalence of VHDL in real-world designs, there is no benchmark evaluating LLMs at the scale and complexity of entire VHDL repositories. Existing VHDL benchmarks, such as VHDL-Eval, focus on small, isolated modules and do not assess model capabilities on project-wide type definitions, shared packages, architectural variations, or hierarchical design. By contrast, recent Verilog benchmarks like RTL-Repo~\cite{allam2024rtl} highlight the importance of long-context modeling and cross-file dependency reasoning, but they are Verilog-specific and do not capture VHDL's unique semantics, compilation constraints, and hierarchical dependencies. In real VHDL repositories, packages define shared types and functions, architectures instantiate entities from other files, generics propagate constraints, and configuration units bind architectures to entities. Without evaluating these aspects, it is impossible to assess an LLM's readiness for industrial VHDL workflows. Addressing this gap requires a benchmark that preserves full repository granularity rather than reducing VHDL to isolated modules or single files. 

In this work, we present \textsc{Vhdl-RepoBench}, the first large-scale, multi-file, repository-level benchmark specifically for VHDL. We curate $\sim$100 open-source VHDL repositories, encompassing $\sim$2.5k files and $\sim$500 verified testbenches, while maintaining complete project structures including entities, architectures, packages, configuration units, and testbenches. From these repositories, we construct a suite of tasks that reflect realistic RTL design workflows: repository-level line and block completion, specification-to-module generation conditioned on full project context, repository-aware natural-language (NL) summarization, and long-context retrieval for context-driven generation (refer Figure~\ref{fig:vhdl_overview}). RTL-Repo uses edit-similarity metrics (Levenshtein distance) without testbench execution, meaning a structurally plausible but functionally broken architecture scores highly. \textsc{Vhdl-RepoBench} evaluates via GHDL compilation and testbench simulation, measuring functional correctness rather than surface similarity. We evaluate multiple state-of-the-art LLMs, including GPT-4o~\cite{chatgpt}, Llama-3-70B~\cite{llama3modelcard}, Qwen2.5-72B~\cite{bai2023qwen}, CodeLlama-70B~\cite{roziere2023code}, and domain-tuned variants, and conduct detailed analysis across all tasks. Our experiments reveal that while current models perform reasonably well on localized syntax and line-level completion, they struggle with multi-file reasoning, hierarchical design understanding, and summarization of repository-level design intent. 

\textbf{Contributions}: Our key contributions are: (a) the introduction of \textsc{Vhdl-RepoBench}, the first large-scale, repository-level VHDL benchmark, containing approximately 2.5k real-world VHDL files with preserved hierarchy and semantics; (b) the development of a multi-task evaluation suite covering Next-Line Prediction (LP), Block Completion (BC), Specification-to-Module Generation (SMG), and Repository Summarization (RS) for VHDL; and (c) comprehensive evaluations of state-of-the-art LLMs, revealing core challenges in VHDL reasoning and highlighting research opportunities for next-generation AI-assisted RTL design tools

\begin{table}[t!]
\caption{Comparison of \textsc{Vhdl-RepoBench} with prior RTL datasets. Dataset size indicates the number of files or problems as reported. Funct. denotes functional verifiability and Lang. indicates the programming language.}
\label{tab:rtl_datasets}
\centering
\small
\begin{tabular}{lcccc}
\toprule
\textbf{Dataset} & \textbf{Size} & \textbf{Funct.} & \textbf{Cross-File}  & \textbf{Lang.}\\
\midrule
\textsc{HumanEval}~\cite{chen2021evaluating} & 264 &  \checkmark & \xmark & Python\\
\textsc{RepoCoder}~\cite{zhang2023repocoder} & 2k & \checkmark & \xmark & Python \\
\textsc{VerilogEval}~\cite{verilogeval} & 156 & \checkmark & \xmark  & Verilog \\
\textsc{Vhdl-Eval}~\cite{vijayaraghavan2024vhdl} & 202 & \checkmark & \xmark & VHDL\\
\textsc{RTLLM}~\cite{rtllm} & 30 & \checkmark & \xmark   & Verilog\\
\textsc{RTLRewriter}~\cite{rtlrewriter} & 95 & \xmark & \checkmark  & Verilog \\

\textsc{CVDP}~\cite{pinckney2025comprehensive} & 783 & \checkmark & \checkmark & Verilog   \\
\textsc{ResBench}~\cite{guo2025resbench} & 56 & \checkmark & \checkmark & Verilog \\
\textsc{TuRTLe}~\cite{garcia2025turtle} & 1k & \checkmark & \checkmark & Verilog  \\
\textsc{RTL-Repo}~\cite{allam2024rtl} & 4k & \xmark & \checkmark & Verilog\\
\midrule
\textbf{\textsc{Vhdl-RepoBench }} & \textbf{2.5k} & \textbf{\checkmark} & \textbf{\checkmark} & \textbf{VHDL} \\
\bottomrule
\end{tabular}

\end{table}

\section{Related Work}
\label{sec:related}

Benchmarking large language models (LLMs) for code generation and understanding has progressed rapidly, beginning with software-oriented datasets. Early benchmarks like \textsc{HumanEval}~\cite{chen2021evaluating} evaluate short, self-contained Python functions, while \textsc{RepoCoder}~\cite{zhang2023repocoder} extends evaluation to the repository-level. Although these datasets have shaped LLM evaluation methodology, they do not reflect hardware design constraints, particularly the concurrency, timing semantics, and cross-file organization inherent to RTL codebases.

Within HDLs, several benchmarks have emerged that focus on functional correctness at the module level. \textsc{VerilogEval}~\cite{verilogeval} and \textsc{Vhdl-Eval}~\cite{vijayaraghavan2024vhdl} test single-file RTL generation using simulation-based verification. Related datasets such as \textsc{RTLLM}~\cite{rtllm} and \textsc{RTLRewriter}~\cite{rtlrewriter} address syntax correction and rewriting, but similarly limit evaluation to isolated modules and do not incorporate repository-scale structure. As summarized in Table~\ref{tab:rtl_datasets}, these datasets vary modestly in size and are largely confined to short, verifiable problems without multi-file dependencies. More recent progress examines multi-file reasoning and long-context modeling in Verilog. Benchmarks such as \textsc{TuRTLe}~\cite{garcia2025turtle}, \textsc{CVDP}~\cite{pinckney2025comprehensive}, and \textsc{ResBench}~\cite{guo2025resbench} preserve cross-file module interactions, revealing that state-of-the-art LLMs struggle with dependency resolution, interface consistency, and hierarchical understanding in RTL designs. The largest of these, \textsc{RTL-Repo}~\cite{allam2024rtl}, aggregates more than 4k Verilog code samples and enables repository-scale code completion and long-context generation. However, all existing repository-level datasets focus exclusively on Verilog, even though VHDL remains central in high-assurance and FPGA-driven development. As Table~\ref{tab:rtl_datasets} illustrates, no prior benchmark combines large scale, cross-file dependencies, and real-world repository structure for VHDL. This absence is particularly limiting because VHDL exhibits design characteristics not present in Verilog, including strong typing, extensive use of packages, explicit generics, and the entity–architecture separation that distributes semantics across multiple files. Existing VHDL datasets are single-file and therefore cannot assess reasoning across packages, architectures, and configuration units, nor can they evaluate retrieval or summarization at the repository level.

To address this gap, \textsc{Vhdl-RepoBench} introduces the first large-scale, repository-level benchmark dedicated to VHDL. It comprises more than 2.5k real-world VHDL files, preserving complete project hierarchies and cross-file relationships. The benchmark supports tasks such as block and line completion, long-context retrieval, project-level summarization, and specification-to-module generation within authentic repository structure. By placing VHDL on equal footing with previously Verilog-only benchmarks, \textsc{Vhdl-RepoBench} provides the first opportunity to systematically measure LLM capabilities in type-rich, multi-file, production-grade VHDL environments.

\section{Vhdl-RepoBench}
\label{sec:vhdl_repo}

\textsc{Vhdl-RepoBench} is a repository-level VHDL benchmark for evaluating LLMs on realistic RTL workflows (Fig.~\ref{fig:vhdl_overview}). It preserves hierarchy, cross-file dependencies, and verification artifacts, enabling tasks such as repository completion, summarization, and specification-to-module generation.

\subsection{Dataset Construction}

\subsubsection{Repository Collection}

We began by retrieving all public GitHub repositories containing \texttt{.vhdl} or \texttt{.vhd} files, producing an initial pool. To focus on meaningful multi-file designs, we retained only repositories in which at least 70\% of RTL files were VHDL, allowing auxiliary files such as testbenches, documentation, or limited Verilog/SystemVerilog sources. This filtering yielded diverse designs across domains including cryptography (\texttt{AES}, \texttt{PRESENT}), processor cores (\texttt{Neo430}, \texttt{MIPS2}), peripherals (\texttt{LED64}, \texttt{HyperRAM}), and accelerators (\texttt{EdgeDetectionAccelerator}). Only permissively licensed projects (MIT, Apache-2.0, BSD) were kept. We removed duplicates via file hashing and pruned forks or near-identical repositories using tree similarity and manual inspection, ensuring each retained project provides unique design content.

\subsubsection{Repository Filtering}

To guarantee dataset quality and practical relevance, repositories were filtered according to several criteria. Toy repositories, including those with missing architectures or packages, were removed through a combination of automated heuristics and manual curation. Additionally, repositories were screened for testbenches, which are critical for functional correctness evaluation. Autogenerated code, which often occurs in template-based designs or synthesis tool outputs, was also removed to prevent bias in task evaluation. Following this pipeline, the final dataset comprises $\sim$100 repositories with over $\sim2.5k$ VHDL files, representing a wide spectrum of design complexity and application domains.

\begin{figure}[t!]
\centering
\includegraphics[width=0.48\textwidth]{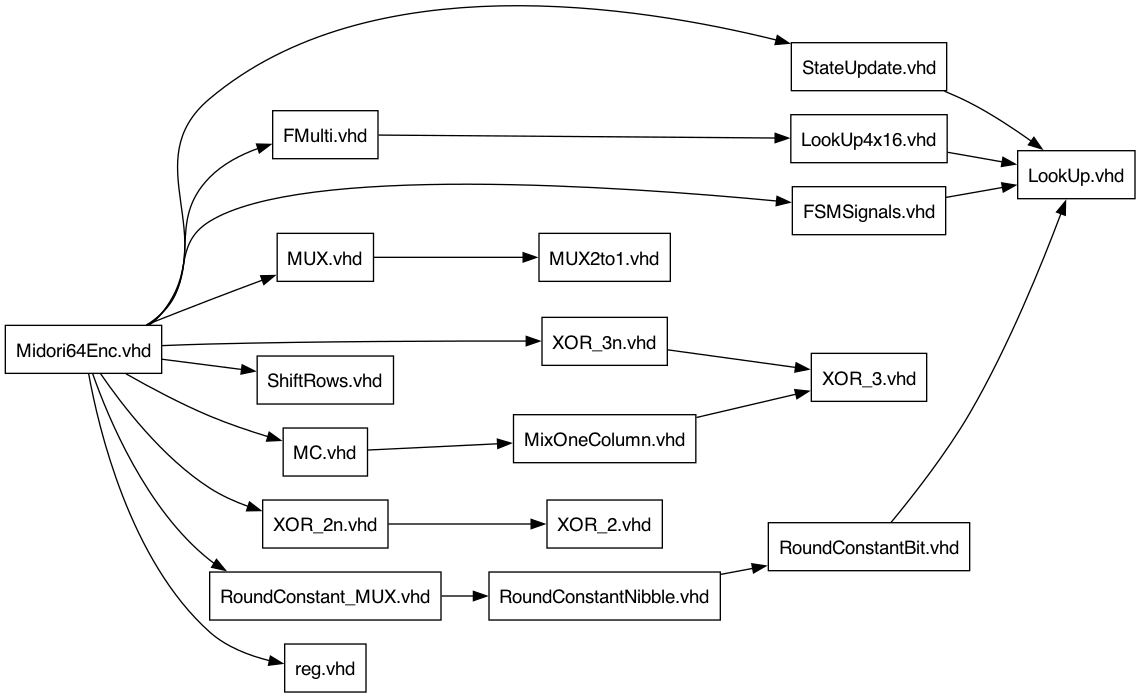}
\caption{Sample cross-file dependency graph for \textit{Midori64} block cipher~\cite{banik2015midori}, where the top-level file \textit{Midori64Enc.vhd} instantiates submodules such as \textit{FMulti.vhd} and \textit{MC.vhd}.}
\label{fig:dependency_graph}
\end{figure}

\subsubsection{Cross-File Dependency Extraction}

Real-world VHDL designs heavily rely on cross-file interactions, such as packages, generics, and configuration units. To systematically capture these relationships, we generate dependency graphs for each repository. Each design is represented as a directed acyclic graph (DAG), where nodes correspond to source files (entities, architectures, or packages) and edges denote instantiation or usage relationships, including: (a) Package imports for shared types, constants, and subprograms; (b) Generic maps connecting entity parameters to architecture implementations; (c) Entity–architecture linking to preserve hierarchical relationships; and (d) Module-level call graphs to track subprogram and procedure usage across files.

These graphs provide a concise view of both hierarchical depth and cross-file breadth, enabling us to filter out trivial or single-file implementations while retaining those with meaningful multi-file structure. Figure~\ref{fig:dependency_graph} illustrates a dependency graph for the \texttt{Midori64} block cipher~\cite{banik2015midori,impeccable_circuits}, where the top-level file \texttt{Midori64Enc.vhd} instantiates submodules present in other files like \texttt{FMulti.vhd} and \texttt{MC.vhd}, which themselves depend on additional files. This captures the hierarchical organization and cross-file dependencies inherent in complex VHDL designs. By using these dependency graphs as a filtering criterion, we ensure that only repositories exhibiting substantial cross-file interactions are retained, providing realistic challenges for hierarchical reasoning and dependency resolution. These retained graphs also serve as the backbone for task scaffolding, supporting repository-level code completion, summarization, retrieval, and specification-to-module generation tasks.

\subsubsection{Dataset Statistics}
Table~\ref{tab:vhdl_dataset_statistics} summarizes key statistics of \textsc{Vhdl-RepoBench}. Repositories exhibit substantial cross-file structure: the average cross-file edge count is 29.6 and average maximum dependency depth is 3.4 levels. All $\sim$1.2k SMG-Module instances include at least one dependency file confirming that tasks genuinely require cross-file reasoning rather than single-file completion.


\begin{table}[b!]
\caption{Key statistics of the \textsc{Vhdl-RepoBench} dataset. Files, entities, architectures, and testbenches are counted across all $\sim$100 repositories.}
\label{tab:vhdl_dataset_statistics}
\centering
\small
\begin{tabular}{lc}
\toprule
\textbf{Statistic} & \textbf{Value} \\
\midrule
\# Repositories & $\sim$100 \\
\# VHDL files & $\sim$2.5k \\
\# Testbenches & $\sim500$ \\
Avg. designs per repository & 1.26 \\
Avg. Cross-file edges per repo & 29.6 \\
Avg. Max dependency depth & 3.4 \\

\bottomrule
\end{tabular}

\end{table}

\begin{figure*}[ht]
    \centering
    \includegraphics[width=0.8\textwidth]{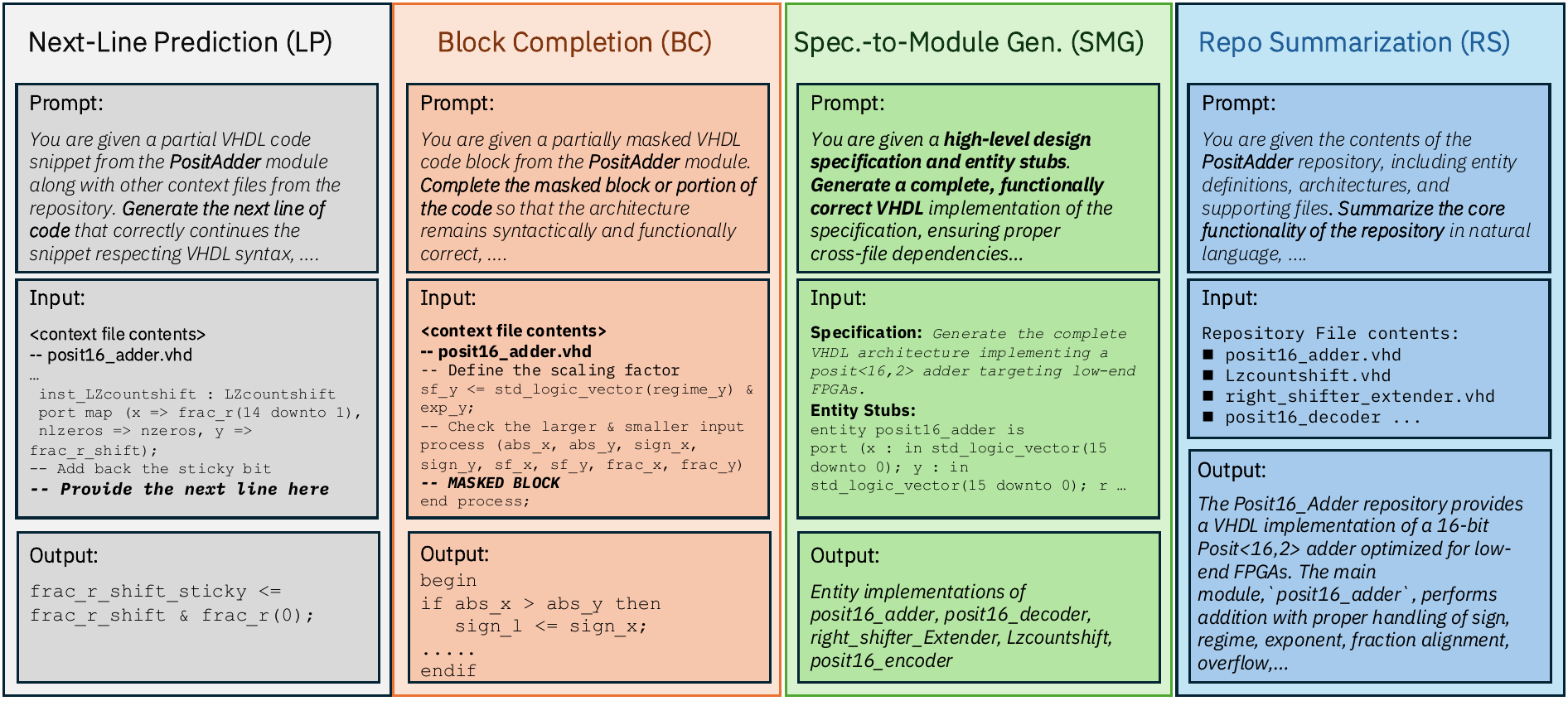}
    \caption{Example task formulations for \textsc{Vhdl-RepoBench} showing the prompt provided to the LLM, input context from the PositAdder repository, and the expected output for each task: Next-Line Prediction (LP), Block Completion (BC), Spec.-to-Module Generation (SMG-R), and Repository Summarization (RS).}
    \label{fig:task_overview}
\end{figure*}

\subsection{Verification and Validation}
\label{sec:verification}

To ensure the correctness and usability of the \textsc{Vhdl-RepoBench} dataset, we implement a rigorous verification and validation pipeline. For each curated repository, we create \texttt{Makefile}s using a combination of human expertise and LLM assistance. These Makefiles execute the available testbenches and confirm both functional correctness and proper hierarchical connectivity across files. This process is essential because the benchmark tasks: Next-Line Prediction (LP), Block Completion (BC), Specification-to-Module Generation (SMG), and Repository Summarization (RS), depend on accurate multi-file relationships, valid VHDL semantics and verified design functionality. The repositories in the dataset exhibit substantial heterogeneity, including variations in directory layout, file ordering, and testbench type. Some repositories include self-verifying testbenches that automatically check expected behavior for some test cases, while others only compile and run, ensuring proper cross-file dependency resolution. To standardize evaluation across this diversity, the \texttt{Makefile}s encode the correct simulation workflow for each repository, abstracting differences in testbench frameworks such as \texttt{GHDL}, \texttt{VUnit}, or \texttt{Cocotb}~\cite{rosser2018cocotb}. This ensures that every design can be evaluated via a single, uniform \texttt{make} command, with the \texttt{Makefile} internally managing framework-specific and repository-specific compilation and execution steps. By employing this mechanism, we achieve: (a) {Functional correctness}: each repository can be compiled and executed to verify that the code behaves as intended; (b) {Hierarchical and cross-file validation}: entity-architecture links, package imports, and submodule instantiations are preserved; (c) {Reproducibility}: all testbenches can be consistently executed across different repositories using a unified interface; and (d) {Dataset quality assurance}: invalid, incomplete, or broken repositories are identified and filtered, guaranteeing that downstream tasks (LP, BC, SMG, RS) are evaluated on reliable data. This verification step distinguishes \textsc{Vhdl-RepoBench} from larger datasets like RTL-Repo, which rely on edit similarity and exact match without testbench execution, potentially reducing quality. By enforcing functional execution and cross-file correctness, our dataset provides a more reliable benchmark for VHDL code understanding and generation.


\begin{table*}[ht]
\caption{Summary of tasks in \textsc{Vhdl-RepoBench}, including approximate size, input, and output format. Here, SMG refers to Specification-to-Module Generation (SMG)}
\label{tab:task_summary}
\centering
\small
\begin{tabular}{lccc}
\toprule
\textbf{Task} & \textbf{Size (approx.)} & \textbf{Input} & \textbf{Output} \\
\midrule
Next-Line Prediction (LP) & $\sim7$k & Preceding lines (local/cross-file context) & Next line of code \\
Block Completion (BC) & $\sim1$k & Partial file or masked block & Completed block/file \\
Repository Summarization (RS) & 850 & Repository files & NL summary \\
SMG-Module (SMG-M) & $\sim$1.2k & Entity stub + Specification + Cross-file Context & Architecture Body\\
SMG-Repo (SMG-R) & 20 & System-Level Spec + Entity Signatures & Complete RTL source \\
\bottomrule
\end{tabular}

\end{table*}

\subsection{Task Formulation}
\label{sec:tasks}
\noindent To evaluate LLM capabilities on multi-file VHDL repositories, we define four tasks that capture both localized code generation and repository-level reasoning. Each targets key aspects of VHDL understanding, including syntax, semantics, hierarchy, and design intent. The tasks are as follows:

    \noindent \textbf{Next-Line Prediction (LP)}: Predict the next line of VHDL code given preceding context within a file or across files. 

    \noindent \textbf{Block Completion (BC)}: Generate larger coherent code regions, including masked code blocks, entire processes, or complete files. 

    \noindent \textbf{Repository Summarization (RS)}: Produce structured NL summaries of a repository’s functionality, hierarchy, and interconnections across its entities, architectures, and packages.

    \noindent \textbf{Specification-to-Module Generation (SMG)}: 
    Generate functional VHDL from specifications and repository context. \textit{SMG-Module}  (SMG-M) generates entity architectures, while \textit{SMG-Repo} (SMG-R) generates complete RTL repositories from system specifications. Dependency graphs preserve cross-file hierarchy and interactions. All datasets are curated from VHDL repositories and quality-verified.

\subsubsection{Next-Line Prediction (LP)}
The Next-Line Prediction task evaluates the model's ability to generate the subsequent line of VHDL code given preceding lines. It measures immediate syntactic correctness, type consistency, and local semantic validity in both local-file and cross-file contexts.

    \noindent \textbf{Input:} VHDL snippet containing all preceding lines within a process, architecture, or entity. Context can include relevant lines from other files in the repository (cross-file).  
    
    \noindent \textbf{Output:} The predicted next line that correctly continues the snippet, respecting syntax, types, and dependencies.

\noindent \textbf{Data Preparation:} Examples are sampled from architecture and process blocks across curated repositories. To focus on meaningful lines, we exclude structural keywords (e.g., \texttt{begin}, \texttt{end}), comments, and simple signal declarations, selecting lines that contribute substantive computation or control flow. Autogenerated or trivial lines are further filtered using file-hash deduplication. A subsample of the dataset is qualitatively verified to ensure high quality.

\noindent \textbf{Metrics:} Exact match accuracy and Pass@k based on verification success (see Sec.~\ref{sec:verification}).

\subsubsection{Block Completion (BC)}
Block Completion evaluates hierarchical and cross-file reasoning by generating contiguous regions of code. This task assesses both local coherence and global repository-level consistency, requiring models to respect inter-file dependencies and architectural hierarchy.

    \noindent \textbf{Input:} An architecture, entity, or file with masked code blocks.  
    
    \noindent\textbf{Output:} Completed code blocks that correctly integrate with both local context and cross-file dependencies, including packages, generics, and instantiations of other entities.  

\noindent \textbf{Data Preparation:} 
Candidate blocks are selected using repository dependency graphs, prioritizing regions whose removal breaks verification or synthesis. Masked blocks are validated through syntax checks and self-verifying testbenches with dependent files. Cross-file context includes relevant packages, entity interfaces, and generics from repository relationships rather than arbitrary file concatenation.

\noindent \textbf{Metrics:} BC is evaluated using Pass@1 and Pass@3 based on GHDL compilation success.

\subsubsection{Repository Summarization (RS)}
Repository Summarization (RS) evaluates a model's ability to generate structured, high-level descriptions of a VHDL repository by capturing its functionality and hierarchical organization. Unlike code-generation tasks, RS focuses on design-intent understanding and is therefore excluded from Pass@k compilation evaluation. Its primary purpose is to assess whether models can reason about and explain repository-level structure, a foundational capability for specification-driven hardware design.


\noindent \textbf{Input:} Complete repository files including entities, and packages, preserving cross-file dependencies.  

    \noindent \textbf{Output:} A coherent NL summary that explains the functionality, design intent of modules/files within the repository.  

\noindent \textbf{Data Preparation:} Summaries and specifications are initially generated using LLMs (e.g., DeepSeekV3~\cite{liu2024deepseek}) from repository code, comments, and documentation. To ensure quality and fidelity, about 10\% of repositories are manually audited, confirming that the summaries accurately capture hierarchical structure, module interactions, and design intent.


\noindent \textbf{Metrics:} Summarization quality is evaluated using ROUGE-L for lexical overlap and structural alignment, along with a \emph{Preferred Rate (PR)}, where an LLM acts as a judge to assess correctness of the summary relative to the repository context.

\subsubsection{Specification-to-Module Generation (SMG)}
SMG evaluates a model's ability to generate functional VHDL architectures from high-level design specifications, correctly resolving cross-file dependencies and integrating with existing modules and packages. We provide two variants spanning entity-level and system-level generation.

\noindent \textbf{SMG-Module} targets one VHDL entity per instance. The model receives an entity stub (ports, generics), a natural-language specification, and cross-file context assembled from the repository dependency graph, and must generate the complete architecture body. This is the primary evaluation target for statistical comparisons.

\noindent \textbf{SMG-Repo} targets end-to-end system design. The model receives a system-level functional specification, top-level entity signature, and package stubs, and must generate \emph{complete} RTL source files.

\noindent \textbf{Data Preparation:} NL specifications are generated by DeepSeekV3~\cite{liu2024deepseek} from repository code and documentation; $\sim$10\% of instances are manually audited for fidelity. All instances are verified by GHDL compilation. The 135 SMG-Module instances with testbenches are additionally validated by testbench execution, confirming functional correctness and cross-file dependency preservation. Grounding specifications in real repository structure mitigates LLM self-alignment bias noted in prior work.

\noindent \textbf{Metrics:} Pass@1 via the unbiased estimator~\cite{chen2021evaluating}, based on GHDL compilation success; testbench execution is additionally required for the verified subset.


    



\begin{table*}[ht]
\caption{Performance of evaluated LLMs on \textsc{Vhdl-RepoBench}. LP: Next-Line Prediction, BC: Block Completion, SMG: Specification-to-Module Generation, RS: Repository Summarization. Context indicates maximum tokens the model can process. CoDes ~\cite{vijayaraghavan2024chain} and Reflexion~\cite{Reflexion2023} show competitive improvements over base Qwen2.5-72B.}
\label{tab:results}
\centering
\small
\begin{tabular}{lcccccccc}
\toprule
\textbf{Model} & \textbf{Context (tokens)} & \multicolumn{2}{c}{\textbf{LP}} & \textbf{BC (Pass@1)} & \multicolumn{2}{c}{\textbf{SMG (Pass@1)}} & \multicolumn{2}{c}{\textbf{RS}} \\
\cmidrule(lr){3-4} \cmidrule(lr){6-7} \cmidrule(lr){8-9}
 & & Exact Match & Pass@1 &  & SMG-R & SMG-M & ROUGE-L & PR \\

\midrule
GPT-4o              & 128K & 59.2 & 60.9 & 45.1 & 19.8 & 41.3 & 44.5 & 0.49 \\
Qwen2.5-72B         & 128K & 48.7 & 53.1 & 38.2 & 15.0 & 34.8 & 37.6 & 0.40 \\
Llama-3-70B         & 128K & 48.3 & 48.9 & 36.9 & 14.2 & 33.1 & 38.4 & 0.41 \\
CodeLlama-70B       & 16K  & 39.2 & 40.3 & 29.8 & 10.2 & 21.4 & 28.7 & 0.26 \\
Verigen             & 2K   & 26.4 & 24.8 & 17.5 &  6.5 & 12.8 &  -   &  -   \\
\midrule
Qwen2.5-72B + Reflexion & 128K & 56.2 & 56.7 & 41.7 & 16.5 & 37.2 & 39.5 & 0.43 \\
Qwen2.5-72B + CoDes     & 128K & 55.4 & 57.2 & 40.4 & 16.8 & 36.9 & 38.6 & 0.41 \\
\bottomrule

\end{tabular}

\end{table*}

\section{Experiments}
\label{sec:experiments}

We evaluate several LLMs on \textsc{Vhdl-RepoBench} to assess VHDL code generation and summarization. The evaluation suite consists of the four tasks in Section~\ref{sec:tasks}: LP, BC, SMG, and RS. Evaluated models include GPT-4o, Verigen, CodeLlama-70B, and Qwen2.5-72B. We also study the effect of structured multi-step reasoning using Reflexion ~\cite{Reflexion2023} and CoDes~\cite{vijayaraghavan2024chain} applied to Qwen2.5-72B.


\textbf{Data Preparation.} LP and BC examples are sampled from curated repositories using dependency graphs to ensure that selected lines or blocks are semantically meaningful and would break compilation or testbench execution if removed. SMG pairs entity stubs with generated specifications, and correctness is validated through compilation and execution of self-verifying testbenches. RS prompts are produced using LLMs (CodeLlama-70B~\cite{roziere2023code} and GPT-4o~\cite{chatgpt}) and manually checked on approximately ten percent of repositories to ensure high-quality summaries. For all tasks, models are prompted in a standardized setting. Code-generation tasks use sampling with temperature $0.2$, top-$k$ of $50$, and top-$p$ of $1.0$. Context limits follow model constraints: approximately $128$K tokens for GPT-4o, Llama-3-70B, and Qwen2.5-72B; $16$K for CodeLlama-70B; and $2$K for Verigen. Limited context windows disadvantage certain models on large repositories, reflecting practical deployment constraints. To benchmark multi-step reasoning, we employ \textit{Reflexion}~\cite{Reflexion2023} and \textit{CoDes}~\cite{vijayaraghavan2024chain}, which generate intermediate reasoning steps. This enables systematic evaluation of complex VHDL tasks and a model's ability to decompose problems. Code generation tasks use standardized sampling (temperature $0.2$, top-$k=50$, top-$p=1.0$) and context limits of $\sim128$K tokens (GPT-4o, Llama-3-70B, Qwen2.5-72B), $16$K (CodeLlama-70B), and $2$K (Verigen). Limited context windows may disadvantage models on large repositories. Since data is sourced from public GitHub repositories, training contamination is possible. Despite deduplication, overlap with pretraining corpora cannot be fully ruled out, affecting LP and BC interpretation.


\textbf{Evaluation Metrics.} LP is evaluated using Exact Match and Pass@1 (compilation success)~\cite{chen2021evaluating}. For SMG and BC, Pass@1 requires GHDL compilation, and for a verified subset, passing testbench execution. RS is evaluated using ROUGE-L and Preferred Rate (PR). Unlike RTL-Repo's Levenshtein edit distance, which measures surface similarity, our Pass@k metrics assess compilability and functionally correctness.

\section{Results}
\label{sec:results}

Table~\ref{tab:results} shows performance across all four \textsc{Vhdl-RepoBench} tasks. Three trends emerge: (1) large-context models achieve higher repository-level accuracy, (2) GPT-4o is the strongest overall, and (3) multi-step prompting improves Qwen2.5-72B across tasks.

\noindent \textbf{Impact of Context Length}

\noindent  Models with 128K-token context windows (GPT-4o, Qwen2.5-72B, Llama-3-70B) consistently outperform smaller-context models. Long-context access is essential for VHDL, where correctness frequently depends on type declarations, package imports, and hierarchical instantiations distributed across multiple files. In contrast, CodeLlama-70B (16K) and Verigen (2K) exhibit sharp performance degradation, particularly in BC and SMG, where missing cross-file information directly limits correctness. This confirms that repository-scale HDL tasks fundamentally require long-context modeling.

\noindent \textbf{Impact of Structured Multi-Step Reasoning}

\noindent Applying Reflexion~\cite{Reflexion2023} and CoDes~\cite{vijayaraghavan2024chain} to Qwen2.5-72B yields meaningful improvements across LP, BC, and SMG. Reflexion helps reduce syntactic inconsistencies through iterative self-critique, while CoDes decomposes repository-level structure into intermediate descriptions before generation. Although neither surpasses GPT-4o, both significantly narrow the gap relative to the base Qwen2.5-72B model, demonstrating the value of structured reasoning.

\noindent \textbf{Task-Wise Analysis}

\noindent  \textit{Next-Line Prediction (LP).}
LP evaluates fine-grained code understanding. GPT-4o achieves the best performance (60.9 Pass@1), with Qwen2.5-72B and Llama-3-70B following closely. CodeLlama-70B \& Verigen struggle due to truncated context and weaker VHDL-specific priors. Reflexion and CoDes boost Qwen2.5-72B by improving line-level consistency and reducing hallucinated identifiers.

\noindent \textit{Block Completion (BC).}
BC requires generating syntactically and semantically coherent multi-line segments. GPT-4o leads with 45.1 Pass@1, benefiting from long-range dependency tracking. Context-limited models degrade substantially, confirming the necessity of cross-file visibility. Reflexion provides the largest improvement for Qwen2.5-72B on this task by enabling iterative structural correction.

 \noindent \textit{Specification-to-Module Generation (SMG).}
SMG is the most challenging task. For SMG-Repo, which requires generating all RTL files from a system-level specification, GPT-4o achieves Pass@1 of 19.8, with long-context models substantially outperforming short-context ones. For SMG-Module (entity-level generation across $~\sim1.2k$ instances), GPT-4o reaches 41.3, confirming that entity-level generation is meaningfully more tractable than full-repository synthesis. Reflexion and CoDes improve Qwen2.5-72B on both variants by decomposing repository structure before generation.

\noindent \textit{Repository Summarization (RS).}
RS evaluates a model’s ability to explain design hierarchy and module intent. GPT-4o achieves the highest ROUGE-L and PR. Qwen2.5-72B and Llama-3-70B perform well with full context, while CodeLlama-70B underperforms due to a limited window. Reflexion and CoDes give modest but consistent gains by promoting structured decomposition. 

\noindent \textit{Error Analysis.}
Three recurring failures appear: (i) incorrect cross-file references (missing imports/entity mismatches), (ii) structurally correct but functionally failing modules, and (iii) degraded performance on long dependency chains, highlighting limits in hierarchical reasoning and global consistency.

\section{Conclusion}
\label{sec:conclusion}
We introduce \textsc{Vhdl-RepoBench}, the first benchmark for evaluating LLMs on repository-scale VHDL understanding, generation, and cross-file reasoning. Tasks mirror realistic workflows: next-line prediction, block-level completion, specification-driven module or repository generation, and repository summarization, capturing challenges absent in file-level HDL datasets. Long-context models (GPT-4o, Qwen2.5-72B, Llama-3-70B) outperform shorter-context models, showing the value of repository context for hierarchical and cross-file reasoning. GPT-4o achieves the highest accuracy, while multi-step prompting strategies like Reflexion and CoDes boost Qwen2.5-72B. Yet, performance on specification-to-module generation and block-level reconstruction remains insufficient for reliable RTL automation. \textsc{Vhdl-RepoBench} lays a foundation for evaluating EDA-focused LLMs and highlights directions for future research: better structural reasoning, broader context integration, and hardware-aware constraints. 
\bibliographystyle{ieeetr}
\bibliography{main}

\end{document}